\documentclass[prc,aps,preprint,showpacs]{revtex4}
\usepackage{graphicx}

\begin{document}
\title{Nilsson diagrams for light neutron-rich nuclei with weakly-bound 
neutrons} 

\author{ Ikuko Hamamoto }

\affiliation{
{\it Division of Mathematical Physics, Lund Institute of Technology 
at the University of Lund, Lund, Sweden}   
}




\begin{abstract}
Using Woods-Saxon
potentials and the eigenphase formalism for one-particle resonances, 
one-particle bound and resonant levels for neutrons 
as a function of quadrupole deformation 
are presented, which are supposed  
to be useful for the interpretation of spectroscopic properties of some 
light neutron-rich nuclei with weakly-bound neutrons.  
Compared with Nilsson diagrams in text books which are constructed 
using modified
oscillator potentials, we point out a systematic change of the shell structure 
in connection with both 
weakly-bound and resonant one-particle levels related to small 
orbital angular momenta $\ell$.
Then, it is seen 
that weakly-bound neutrons in nuclei such as $^{15-19}$C and $^{33-37}$Mg 
may prefer to being deformed as a result of Jahn-Teller effect, due to the near
degeneracy of the 1d$_{5/2}$-2s$_{1/2}$ levels and the 1f$_{7/2}$-2p$_{3/2}$ 
levels in the spherical potential, respectively.  
Furthermore, the absence of some 
one-particle resonant levels compared
with the Nilsson diagrams in text books is illustrated.
\end{abstract}


\maketitle

\section{INTRODUCTION}
The study of one-particle motion in spheroidal potentials, which is the basis
for the understanding of deformed nuclei, started 
in the fifties \cite{HW53,SM55,SGN55,KG56}.  In particular, the work by S. G.
Nilsson \cite{SGN55} played an important role for years in providing the 
basis for the classification of experimental data on the spectra of stable 
odd-A deformed nuclei \cite{BM75}.  Since the nucleon separation energy in
stable nuclei is 7-10 MeV, the spectroscopic analysis around the ground
state of those nuclei 
has been successfully performed in terms of harmonic-oscillator wave-functions.
In contrast, the recent study of nuclear structure close to the neutron drip
line points out the unique role of weakly-bound neutrons with small angular
momentum $\ell$ and the importance of coupling to the nearby continuum of
unbound states.  
Due to the absence of the Coulomb barrier, weakly-bound neutrons with small
$\ell$ have an appreciable probability to be outside the core nucleus.  
Thus, those neutrons are insensitive to the strength of the potential provided
by the well-bound nucleons in the system.  In particular, the behavior of
s$_{1/2}$ neutrons is an extreme example due to the absence of the centrifugal
barrier for the $\ell$=0 orbit, compared with weakly-bound large $\ell$
neutrons, of which wave functions stay mostly inside the potential.  The $\ell
(\ell +1)$ dependence of the height of the centrifugal barrier affects sharply
the presence (or absence) of one-particle resonant levels at a given positive
energy.  Taking spheroidal Woods-Saxon potentials, the properties of both
weakly-bound neutron orbits and one-neutron resonant levels are 
studied in Refs. \cite{IH04,IH05,IH06}.

There have been a number of self-consistent Hartree-Fock (HF)
calculations of light neutron-rich nuclei, 
in many of
which deformed HF solutions are indeed obtained. 
(As a classic work we may mention Ref. \cite{CFKK75}.)
However, almost all those
calculations are done either using the expansion in terms 
of harmonic oscillator bases 
or confining the system in a finite box.
To our knowledge, no deformed HF
calculation is yet available, which is carried out by integrating 
in mesh of space coordinate with proper
asymptotic behavior for $r = R_{max}$, at which the nuclear potential is
negligible. 
If deformed HF calculations are carried out in the latter way, 
the numerical results
are totally independent of the values of $R_{max}$ and, furthermore, 
it becomes possible to estimate one-particle
resonant levels without any ambiguity.

The effective interactions to be used in HF calculations of nuclei far
away from the stability line are not yet properly fixed.
Thus, if the interaction chosen is not appropriate for those nuclei away from 
the stability line, the answer is not reliable.  
Moreover, one HF solution gives one self-consistent deformation, which is 
determined by the two-body interaction selected, assuming that all technical
problems in obtaining HF solutions are solved.  Thus, it is not easy either 
to pin down the origin of the deformation obtained or 
to evaluate the ambiguity coming from the choice of the two-body interaction.

In the analysis of observed spectroscopic properties of light neutron-rich
nuclei with weakly-bound neutrons the shell model is so far used in most cases. 
The shell model should be applicable to those nuclei, if the configuration 
space is sufficiently
large and the weakly-bound particles are properly treated.  The latter condition
is usually not satisfied, since harmonic-oscillator 
wave-functions in a limited space are used in most cases. 
The work summarized in Ref. \cite{VZ06} is an exception among the calculations 
that can be systematically compared with
experiments, however, the
complicated calculations have so far been carried out only for the helium and
oxygen isotopes.
In any case, some physically interesting quantities 
such as one-particle energies
(or shell structure) or nuclear shape (or deformation) are not directly obtained
from shell-model calculations.

Recognizing how useful it was to have Nilsson diagrams in the study of stable
deformed nuclei \cite{BM75}, 
which are constructed using modified oscillator potentials, in this paper  
we present some "Nilsson diagrams" that are relevant
especially to  
some light neutron-rich nuclei towards the neutron-drip-line. Taking the
Woods-Saxon potentials with the parameters adjusted to 
some particular nuclei, 
both bound and resonant one-neutron levels are calculated as a
function of quadrupole deformation.   
The change of nuclear shell-structure for neutrons 
is seen in both negative and positive 
one-particle energies of the Nilsson diagrams. 
The change comes from the unique behavior of neutron orbits 
with small $\ell$ values, in particular $\ell$=0 and 1.
The modified shell-structure has direct relevance to the ground and 
low-lying states of
neutron-drip-line nuclei, in which weakly-bound neutrons are present.
Considering the possible absence of many-body pair-field in light nuclei, 
the study of the present type of Nilsson diagrams can
definitely help us to understand the origin of possible deformation and the
related spectroscopic properties of light neutron-drip-line nuclei.

In Sect. II some points of our model are summarized, while numerical
results are presented in Sect. III.  Conclusions and discussions are given in
Sect. IV.

\section{MODEL}
The occupancy of weakly-bound one-particle 
levels has a contribution especially to the tail of 
the self-consistent potentials.  However, 
even for light nuclei presently considered the number of weakly-bound
neutron(s) is much smaller than that of well-bound core nucleons.  
Namely, the major part of the nuclear potential is provided by well-bound
nucleons.
Thus, for simplicity, 
the parameters of Woods-Saxon potentials are taken from the standard ones
\cite{BM69} for stable nuclei except for the depth, $V_{WS}$. 
Namely, the diffuseness, the strength of spin-orbit potentials and the radius
parameter are taken from those on p.239 of Ref. \cite{BM69}. 
The depth is adjusted so that a particular one-neutron level obtains a given
desirable binding-energy in respective examples.

The way in which bound one-particle levels are calculated is described in Ref. 
\cite{IH04}, while the eigenphase formalism that is used to estimate
one-particle resonant levels for a deformed potential is given in 
Refs. \cite{IH05,IH06}.
The essential point is : the coupled equations obtained from the 
Schr\"{o}dinger equation are 
solved in coordinate space with the correct asymptotic behavior of wave
functions for $r \rightarrow \infty$.  
The solution obtained in this way is totally independent of the upper limit of
radial integration, $R_{max}$, if both the potential and the coupling term 
are already negligible at $r = R_{max}$.
One-particle resonant energy for $\beta \neq 0$ is defined 
as the energy, at which one of eigenphases  
increases through $\pi /2$ as energy increases. In the limit of $\beta
\rightarrow 0$ this definition in the eigenphase formalism 
is in agreement with the definition of one-particle resonance in spherical
potentials described in text books \cite{RGN66}; 
the phase shift increases through $\pi$/2 as energy
increases.

One-particle resonance is not obtained if none of calculated 
eigenphases do not increase through $\pi /2$ as energy increases. 
For example, we have no corresponding resonance in the case that a calculated 
eigenphase starts to decrease before reaching $\pi /2$ as energy increases.
Even if one fails to obtain one-particle resonance defined 
in terms of eigenphase, for a certain small region of energy just 
after the disappearance of resonance the concentration of the wave functions
inside the potential may still be found.  
However, the concentration will easily disappear after a short time  
if a resonance is no longer obtained in the eigenphase formalism. 
This situation is analogous to the case of the spherical potential, in which 
the phase shift starts to decrease before reaching $\pi$/2 as energy increases
\cite{RGN66}. 

Compared with the Nilsson diagram based on modified oscillator potentials,
the striking difference of the level scheme obtained in the present work 
comes from the behavior of levels with low $\ell$ values 
(in particular, $\ell$=0 and 1) for $\beta$=0 and those with 
small $\Omega$ values (mainly $\Omega ^{\pi}$ = 1/2$^{+}$, 
1/2$^{-}$ and 3/2$^{-}$) for $\beta \neq 0$, 
in both the weakly-bound and positive-energy region. 
Note that the minimum $\ell$ value of possible components of $\Omega^{\pi}$ = 
1/2$^+$, 1/2$^-$ and 3/2$^-$ levels is equal to 0, 1 and 1, respectively.
The absence of the centrifugal barrier for the $\ell$=0 channel produces 
the unique behavior
of weakly-bound and positive-energy $\Omega^{\pi}$ = 1/2$^+$ orbits.
However, we find that some $\Omega^{\pi} = 1/2^{+}$ resonant levels survive 
in a higher-energy region (see, for example, the [200 1/2] level in Fig. 1), 
if the relative probability of s$_{1/2}$ component inside the potential is
smaller than a certain critical value \cite{IH06}.
Since the height of the centrifugal barrier becomes lower for a larger 
nuclear radius, the unique behavior of $\ell$=1 components will be more easily 
seen in nuclei with larger mass.

\section{NUMERICAL RESULTS}
\subsection{Neutron-rich C-isotopes}
Taking $V_{WS} = -$40.0 MeV and the radius parameter for A=17, 
at $\beta = 0$ in Fig. 1 we obtain  
$\varepsilon (1d_{5/2}) = -560$ keV and 
$\varepsilon$(2s$_{1/2}$) = $-415$ keV.
The resonant energy of the 1d$_{3/2}$ level is 5.60 MeV.
The near degeneracy of the 1d$_{5/2}$ and 2s$_{1/2}$ levels  
compared to the high-lying 1d$_{3/2}$ level exhibits that 
for the spherical shape 
the neutron number N=16 may behave like a magic number.  
For $\beta \neq 0$ 
the s$_{1/2}$, d$_{3/2}$, d$_{5/2}$, g$_{7/2}$ and g$_{9/2}$ 
channels are included in the calculation of positive-parity levels, 
while the p$_{1/2}$,
p$_{3/2}$, f$_{5/2}$ and f$_{7/2}$ channels for negative-parity levels. 
The parameters are chosen approximately for neutrons in the nucleus
$^{17}_{6}$C$_{11}$, since 
the observed neutron separation energy
of $^{17}$C is $-$730 keV and the nucleus is presumably prolately
deformed.  
Examining the Nilsson diagram in Fig. 1, it is easily imagined that a
few neutrons occupying 
the weakly-bound almost-degenerate 1d$_{5/2}$ - 2s$_{1/2}$ shells 
at $\beta = 0$ may 
prefer to being deformed in order to gain the total 
energy (Jahn-Teller effect).  
This may be the case of neutrons in 
$^{15,17,19}$C, of which the neutron separation energy is small.
It is also noticed that the observed ground states of 
$^{17}_{6}$C$_{11}$ and 
$^{19}_{6}$C$_{13}$ with $I^{\pi} =$ 3/2$^{+}$ and 1/2$^{+}$, 
respectively, may be in a natural
way interpreted as the band-heads of the intrinsic 
[211 3/2] and [211 1/2] configurations 
for prolate deformation, when the level scheme in the Nilsson diagram is
applied to those nuclei.  For some experimental evidence for the
deformation of those C-isotopes, see for example \cite{ZE04,ZE05}.

The [220 1/2] resonant level is not obtained for $\beta < -0.12$ and
$\varepsilon_{\Omega} > 0.22$ MeV, because 
the predominant component of the [220 1/2] wave function inside the nuclear
radius is s$_{1/2}$ and thus decays very quickly \cite{IH06}.
Due to the same reason, the continuation of the 
[200 1/2] resonant level cannot be found for 
$\beta > 0.46$ and $0 < \varepsilon_{\Omega} < 1.82$ MeV, while for smaller 
$\beta$ values the predominant component of the [200 1/2] level inside the
nuclear radius is 
d$_{3/2}$ and thus the resonant level is sufficiently well defined.

In order to illustrate the near degeneracy of the 2s$_{1/2}$ and 1d$_{5/2}$
levels at $\beta$=0 in Fig. 1, in Fig. 2 the energy eigenvalues of Woods-Saxon
potentials are shown, which are 
obtained by varying the depth while keeping other parameters the same
as in Fig. 1. The 2s$_{1/2}$ level crosses with the 1d$_{5/2}$ level for $\mid
V_{WS} \mid$ slightly smaller than 40 MeV. 
For reference, the value of $V_{WS}$ that is obtained by applying Eq. (2-182) 
of Ref. \cite{BM69} to $^{17}_{6}$C$_{11}$ is $-$41.3 MeV.

If a larger diffuseness, $a > 0.72$ fm, is used in Fig. 1 keeping other
parameters of the Woods-Saxon potential unchanged, for $\beta$=0 the 2s$_{1/2}$
level appears lower than the 1d$_{5/2}$ level.  Nevertheless, in the region of 
an appreciable size 
of deformation the structure of the Nilsson diagram coming from the 2s$_{1/2}$
and 1d$_{5/2}$ levels remains approximately the same.

\subsection{Neutron-rich Mg-isotopes}
Taking $V_{WS} = -$40.0 MeV, in Fig. 3 the Nilsson diagram for neutrons 
is plotted for the
radius appropriate for A=31. 
For the parameters of Fig. 3 the 2s$_{1/2}$ level for $\beta$=0 is well-bound 
and, therefore, 
it lies approximately in the middle of the 2d$_{5/2}$ and 2d$_{3/2}$ levels, 
just as obtained from the level scheme in the modified oscillator
potential \cite{BM75}, but in contrast to the near degeneracy of the
2s$_{1/2}$-1d$_{5/2}$ levels shown in Fig. 1.
For $\beta \neq 0$ the s$_{1/2}$, d$_{3/2}$, d$_{5/2}$, g$_{7/2}$ and g$_{9/2}$ 
channels are included in the calculation of 
positive-parity levels, while the p$_{1/2}$,
p$_{3/2}$, f$_{5/2}$, f$_{7/2}$, h$_{9/2}$ and h$_{11/2}$ channels 
for negative-parity levels. 
The value of $V_{WS}$ is chosen so that the spin-parities of the ground state 
of 
nuclei $^{31}_{12}$Mg$_{19}$ and $^{33}_{12}$Mg$_{21}$, 1/2$^{+}$ 
\cite{GN05} and 
3/2$^{-}$ \cite{GN07}, respectively, are obtained for $\beta \approx 0.5$  
\cite{TM95,HI01,ZE06}, which may be 
identified as the band-heads of the [200 1/2] and [321 3/2] configurations
with observed energies of about $-$2 MeV.  

The 1f$_{7/2}$ resonant level at $\beta$=0 is obtained at 
$\varepsilon_{\Omega}$=1.59 MeV.  In contrast, the 2p$_{3/2}$ and 
2p$_{1/2}$ resonant levels are not obtained, while the trace of the 2p$_{3/2}$ 
resonant level is expected 
to lie below the 1f$_{7/2}$  resonant level (see Fig. 4).  Nevertheless, 
the energy is too high for the existence of 
a resonance with $\ell$=1.
The complicated behavior of the energies of the $\Omega^{\pi}$ = 
1/2$^{-}$ and 3/2$^{-}$ levels 
for $\beta < 0$ and below 2 MeV in Fig. 3 indicates the influence of the 
2p$_{3/2}$ and 2p$_{1/2}$ levels which do not appear as resonant states 
around $\beta$=0.
Or, more exactly speaking, 
taking the $\Omega^{\pi} = 1/2^{-}$ level expressed by 
the dotted curve as an example, the slope in the region of $\beta < -$0.3 
indicates the 2p$_{3/2}$ level at $\beta$=0 lying lower than the 1f$_{7/2}$
level, while the slope for $-0.2 < \beta < -0.14$ suggests the 2p$_{1/2}$ level
at $\beta$=0 lying higher than the 1f$_{7/2}$ level.

In order to locate the trace of the 2p$_{3/2}$ and 2p$_{1/2}$ levels for 
$V_{WS} = -$40 MeV in Fig. 3, in Fig. 4 the energy eigenvalues of Woods-Saxon
potentials are shown 
as a function of $V_{WS}$ while keeping other parameters the same as
in Fig. 3.  The 2p$_{3/2}$ and 2p$_{1/2}$ resonant levels are not obtained for
$\varepsilon > 0.80$ MeV and $\varepsilon > 0.62$ MeV, respectively, while 
the $\ell$=3 one-particle resonant levels continue to be well defined above
$\epsilon$=2 MeV.  From Fig. 4 it is seen that the 2p$_{3/2}$ resonant level
crosses with the 1f$_{7/2}$ resonant level around $V_{WS} = -$42 MeV.
Since in the present parameterization of Woods-Saxon potential the strength of
the spin-orbit potential is proportional to $V_{WS}$, 
it is somewhat misleading to
draw the figure like Fig. 4 for a large variation of $V_{WS}$.  Nevertheless, 
using Fig. 4 it is easy to locate the trace of the 2p$_{3/2}$ resonant level 
for $V_{WS} = -$40 MeV, which is the depth used in Fig. 3.

The fact that the trace of the 2p$_{3/2}$ resonant level is expected to lie 
below the 1f$_{7/2}$ level 
in the positive energy region of Fig. 3 
indicates that N=28 is not a magic number in the example. 
The near degeneracy of the 1f$_{7/2}$ and 2p$_{3/2}$ levels in the
positive-energy region 
is indeed similar to the level scheme of 
almost degenerate 1d$_{5/2}$ and 2s$_{1/2}$ bound levels in Fig. 1.  
This near degeneracy gives certainly the origin of possible 
deformation when a few weakly-bound 
neutrons occupy 
the 1f$_{7/2}$ - 2p$_{3/2}$ shell. 
Namely, the degeneracy can be used to take a particular combination of the
components so as to lower some level energy as deformation sets in 
(Jahn-Teller effect). 
This situation may correspond to neutron-rich Mg isotopes.

In Fig. 3 the level coming from the 2p$_{1/2}$ level is totally missing since it
does not exist as a
resonant level.  The level denoted as [321 1/2] 
is obtained as a resonant level only for $\beta > 0.24$ and 
$\varepsilon_{\Omega} < 1.67$ MeV.  For $\beta < 0.24$  
the energy of the level becomes larger than 1.67 MeV, 
and the width becomes extremely
large because $\ell$=1 is the predominant component of the wave function 
inside the nuclear radius.  Therefore,  the level 
cannot be identified as a resonance.  
The one-particle resonant level obtained for $\beta$=0, which lies next lowest 
to 1f$_{7/2}$, 
is the 1f$_{5/2}$ level 
found at 8.96 MeV that lies outside the range of Fig. 3. 

Taking $V_{WS} = -$40.0 MeV, in Fig. 5 the Nilsson diagram is plotted for the
radius appropriate for A=37. 
The 1f$_{5/2}$ and 2p$_{3/2}$ resonances at $\beta$=0 
are found at 5.22 and 0.018 MeV with the
widths 2.08 and 0.005 MeV, respectively.  
Since $\varepsilon$(1f$_{7/2}$) = $-$0.66 MeV, the distance between the 
1f$_{7/2}$ and 2p$_{3/2}$ levels is 680 keV, which is again 
very small compared with 
the distance obtained in the case that both levels are well-bound.
This near degeneracy of the 1f$_{7/2}$ and 2p$_{3/2}$ levels at $\beta$=0 
suggests that
weakly-bound neutrons in nuclei with N=21-26 may prefer to being deformed.
In the calculation of Fig. 5 for 
$\beta \neq 0$ the s$_{1/2}$, d$_{3/2}$, d$_{5/2}$, g$_{7/2}$ and g$_{9/2}$ 
channels are included in the calculation of positive-parity levels, 
while the p$_{1/2}$,
p$_{3/2}$, f$_{5/2}$, f$_{7/2}$, h$_{9/2}$ and h$_{11/2}$ channels 
for negative-parity levels. 
The value of $V_{WS}$ is chosen so as to simulate the possible 
neutron-drip-line nucleus $^{37}_{12}$Mg$_{25}$, which may have a neutron
separation energy of a few hundreds keV.  

The 2p$_{1/2}$ resonant level is not obtained 
at $\beta$=0, while for $\beta \neq 0$
no $\Omega^{\pi}$ = 1/2$^-$ one-particle level  
connected to the possible 2p$_{1/2}$ level can survive as a resonant level. 
The $\Omega ^{\pi}$ = 3/2$^-$ resonant level 
(denoted by the dashed curve in Fig. 5) 
connected to the 2p$_{3/2}$ level at $\beta$=0 
cannot survive for $\beta > 0.21$ and $\varepsilon_{\Omega} > 1.31$ MeV, 
because the predominant component of the wave function 
inside the nuclear radius 
is $\ell$=1 and the level decays out quickly due to the low centrifugal 
barrier.
The $\Omega ^{\pi}$ = 1/2$^{-}$ resonant level (denoted by the dotted curve in
Fig. 5) coming from the 1f$_{5/2}$ level cannot
survive as a resonance for $\beta > 0.12$ due to the increasing $\ell$=1
component inside the nuclear radius.

In the region of a few MeV excitation-energy of Fig. 5, 
for both spherical and prolate shape we find no well-defined
one-neutron resonant levels except the [303 7/2] level, 
since Nilsson levels expected in the region have either 
$\Omega ^{\pi}$ = 1/2$^{-}$ or 3/2$^{-}$.

It is noted that the energy distance between the 1f$_{7/2}$ and 
2p$_{3/2}$ levels
at $\beta$=0 in the Woods-Saxon potential becomes as large as several MeV 
when both levels are well bound, as known from the presence of the magic number
N=28 in stable nuclei.

\section{CONCLUSIONS AND DISCUSSIONS}
A few examples of Nilsson diagrams with both bound and resonant levels 
are given, the parameters of which are chosen
to be appropriate for some light neutron-rich nuclei with weakly-bound neutrons,
using Woods-Saxon potentials.  The absence of centrifugal barrier (very low
centrifugal barrier) for $\ell$=0 ($\ell$=1) neutrons produces the 2s$_{1/2}$ 
(2p$_{3/2}$) level close to the 1d$_{5/2}$ (1f$_{7/2}$) level, for both
weakly-bound and low-lying resonant neutrons.  This near degeneracy of the
2s$_{1/2}$-1d$_{5/2}$ and 2p$_{3/2}$-1f$_{7/2}$ levels at $\beta$=0 is 
recognized 
as the basic element of producing deformation for some neutron-rich C-Mg
isotopes, in the case that the proton configuration allows the deformation.

One-neutron resonant levels for $\beta \neq 0$ are estimated using the
eigenphase formalism.  The $\Omega^{\pi}$ = 1/2$^{+}$ resonant level can hardly
survive when the predominant component of the wave function inside the potential
is s$_{1/2}$, while the $\Omega^{\pi}$ = 1/2$^-$ and 3/2$^-$ resonant levels are
not obtained if the predominant component has $\ell$=1 and the energy is higher
than 2 MeV in nuclei with A$>$16.
In some nuclei the absence of those $\Omega^{\pi}$ = 1/2$^+$, 1/2$^-$ 
and 3/2$^-$ resonant levels produces a low density of one-neutron resonant
levels in the region of several MeV.  
How much the low density affects the many-body
correlation such as pair correlation in nuclei towards the neutron drip line 
is a future problem to be studied and may be properly studied, only
when the many-body correlation is studied treating the nearby continuum in a
reasonable manner without discretizing the spectra. 
It is noted 
that the neutron one-particle 
levels obtained from Nilsson diagrams for $\beta \neq$0 are 
those counted for band-head states of odd-N nuclei.  
Thus, rotational states, which are
constructed based on those band-head states, should be in principle observed
using a proper experimental method though high-spin states will have 
very narrow widths in the low-energy region.

If one fails to treat properly weakly-bound neutrons or low-energy neutron
resonant levels with small $\ell$, 
the HF 2p$_{3/2}$
(2s$_{1/2}$) level will not come down 
close to the 1f$_{7/2}$ (1d$_{5/2}$) level.  
In any case, if the 1f$_{7/2}$ or 1d$_{5/2}$ level is appreciably isolated, 
in the absence of pair correlation 
it may be possible to obtain an oblate shape as 
the deformation of the system with a few
neutrons in the 1f$_{7/2}$ or 1d$_{5/2}$ shell.  This is because 
a preferred deformation is oblate at the beginning of the shell filling 
if a single j-shell is isolated (for example, see 
\cite{BM60,HMZ91}), while it is prolate if shells with
different j values are nearly degenerate as in 
the harmonic-oscillator potential \cite{BM53}.
It is an interesting question whether any oblate shape is observed around 
the ground state of light neutron-drip-line nuclei of C-Mg isotopes.

\newpage

\noindent
{\bf\large Figure captions}\\
\begin{description}
\item[{\rm Figure 1 :}]
Neutron one-particle levels as a function of quadrupole deformation.
Parameters of the Woods-Saxon potential are designed approximately for 
the nucleus $^{17}$C.
One-particle levels are denoted by the asymptotic quantum numbers 
[N n$_z$ $\Lambda$ $\Omega$].  The $\Omega$-values are denoted for four
positive-parity levels for $\beta < 0$, since it may be difficult to see the
connection to the levels for $\beta > 0$.
One-particle levels appearing at $\beta$=0 are 1p$_{1/2}$, 1d$_{5/2}$, 
2s$_{1/2}$ and 1d$_{3/2}$ levels at $-$6.77, $-$0.56, $-$0.42 and +5.60 MeV,
respectively.  
One-particle levels in the positive-energy region, of which the phase shift 
(one of eigenphases) for $\beta = 0$ ($\beta \neq 0$) 
does not increase through $\pi$/2 as energy increases, are not plotted. 
The neutron numbers 8 and 16, which are obtained by filling in all lower-lying
levels, are indicated with circles.
\end{description}

\begin{description}
\item[{\rm Figure 2 :}]
Neutron one-particle levels as a function of the depth of Woods-Saxon
potential, $V_{WS}$, for $\beta$=0.  All parameters other than $V_{WS}$ are 
the same as in Fig. 1.  Note that $V_{WS} = -$40 MeV is used in Fig. 1.  
The $\ell$=2 
one-particle resonant level continues to be well defined above 
$\varepsilon = 2$ MeV.
\end{description}

\noindent
\begin{description}
\item[{\rm Figure 3 :}]
Neutron one-particle levels as a function of quadrupole deformation.
Parameters of the Woods-Saxon potential are designed approximately for 
the nucleus $^{31}$Mg.
The $\Omega^{\pi}$ = 1/2$^{-}$ levels are denoted by dotted curves, the 
3/2$^{-}$ levels by dashed curves, the 5/2$^{-}$ levels by dot-dashed curves and
the 7/2$^{-}$ levels by dot-dot-dashed curves, while positive-parity levels are 
plotted by solid curves.
One-particle levels appearing at $\beta$=0 are 1d$_{5/2}$, 2s$_{1/2}$, 
1d$_{3/2}$ and 1f$_{7/2}$ levels, from the bottom to the top.  
Neither 2p$_{3/2}$ nor
2p$_{1/2}$ levels are obtained at $\beta = 0$ 
as one-particle resonant levels and, thus, they
are not plotted in the figure.  
The next low-lying one-particle resonant level for $\beta$=0 is 
the 1f$_{5/2}$ level
at 8.96 MeV that lies outside the range of the figure.
See the text for details and the caption to Fig. 1.
\end{description}

\begin{description}
\item[{\rm Figure 4 :}]
Neutron one-particle levels as a function of the depth of Woods-Saxon
potential, $V_{WS}$, for $\beta$=0.  All parameters other than $V_{WS}$ are 
the same as in Fig. 3.  Note that $V_{WS} = -$40 MeV is used in Fig. 3.  
The $\ell$=3 
one-particle resonant levels continue to be well defined above 
$\varepsilon = 2$ MeV, while the 2p$_{3/2}$ and 2p$_{1/2}$ resonant levels
do not survive for $\varepsilon > 0.80$ MeV and $\varepsilon > 0.62$ MeV, 
respectively.
\end{description}

\noindent
\begin{description}
\item[{\rm Figure 5 :}]
Neutron one-particle levels as a function of quadrupole deformation.
Parameters of the Woods-Saxon potential are designed approximately for 
the nucleus $^{37}$Mg.
One-particle levels appearing at $\beta$=0 are 2s$_{1/2}$, 1d$_{3/2}$, 
1f$_{7/2}$, 2p$_{3/2}$ and 1f$_{5/2}$ levels at $-$7.02, $-$5.28, $-$0.66,
+0.018 and +5.22 MeV, respectively.  
The 2p$_{1/2}$ level is not
obtained as one-particle resonant level.  See the caption to Fig. 3 and 
the text for details.
\end{description}


\vspace{2cm}
\begin{thebibliography}{99}
\bibitem{HW53} D. L. Hill and J. A. Wheeler, Phys. Rev. {\bf 89}, 1102 (1953).
\bibitem{SM55} S. A. Moszkowski, Phys. Rev. {\bf 99}, 803 (1955).
\bibitem{SGN55} S. G. Nilsson, Mat. Fys. Medd. Dan. Vid. Selsk. {\bf 29}, no.
16 (1955).
\bibitem{KG56} K. Gottfried, Phys. Rev. {\bf 103}, 1017 (1956).
\bibitem{BM75} A. Bohr and B. R. Mottelson, {\it Nuclear Structure\/} (Benjamin,
Reading, MA, 1975), Vol.II.
\bibitem{IH04} I. Hamamoto, Phys. Rev. C {\bf 69}, 041306 (2004).
\bibitem{IH05} I. Hamamoto, Phys. Rev. C {\bf 72}, 024301 (2005).
\bibitem{IH06} I. Hamamoto, Phys. Rev. C {\bf 73}, 064308 (2006).
\bibitem{CFKK75} X. Campi, H. Flocard, A. K. Kerman and S. Koonin, Nucl. Phys. 
{\bf A251}, 193 (1975).
\bibitem{VZ06} A. Volya and V. Zelevinsky, Phys. Rev. C {\bf 74}, 064314 (2006).
\bibitem{BM69} A. Bohr and B. R. Mottelson, {\it Nuclear Structure\/} (Benjamin,
Reading, MA, 1969), Vol.I.
\bibitem{RGN66} For example, see R. G. Newton, {\it Scattering Theory of Waves
and Particles\/} (McGraw-Hill, New York, 1966). 
\bibitem{ZE04} Z. Elekes et al., Phys. Lett. {\bf B586}, 34 (2004).
\bibitem{ZE05} Z. Elekes et al., Phys. Lett. {\bf B614}, 174 (2005).
\bibitem{GN05} G. Neyens et al., Phys. Rev. Lett. {\bf 94}, 022501 (2005).
\bibitem{GN07} G. Neyens, talk at ECT* workshop, April 2007.
\bibitem{TM95} T. Motobayashi et al., Phys. Lett. {\bf B346}, 9 (1995).
\bibitem{HI01} H. Iwasaki et al., Phys. Lett. {\bf B522}, 227 (2001).
\bibitem{ZE06} Z. Elekes et al., Phys. Rev. C {\bf 73}, 044314 (2006).
\bibitem{BM60} B. R. Mottelson, in Proc. International School of Physics 
``Enrico Fermi'' XV, 1960, edited by G. Racah, p.44.
\bibitem{HMZ91} I. Hamamoto, B. Mottelson, H. Xie and X. Z. Zhang, Zeit. f 
Phys. D {\bf 21}, 163 (1991).
\bibitem{BM53} A. Bohr and B. R. Mottelson, Mat. Fys. Medd. Dan. Vid. Selsk.
{\bf 27}, no. 16 (1953).


\end{thebibliography}
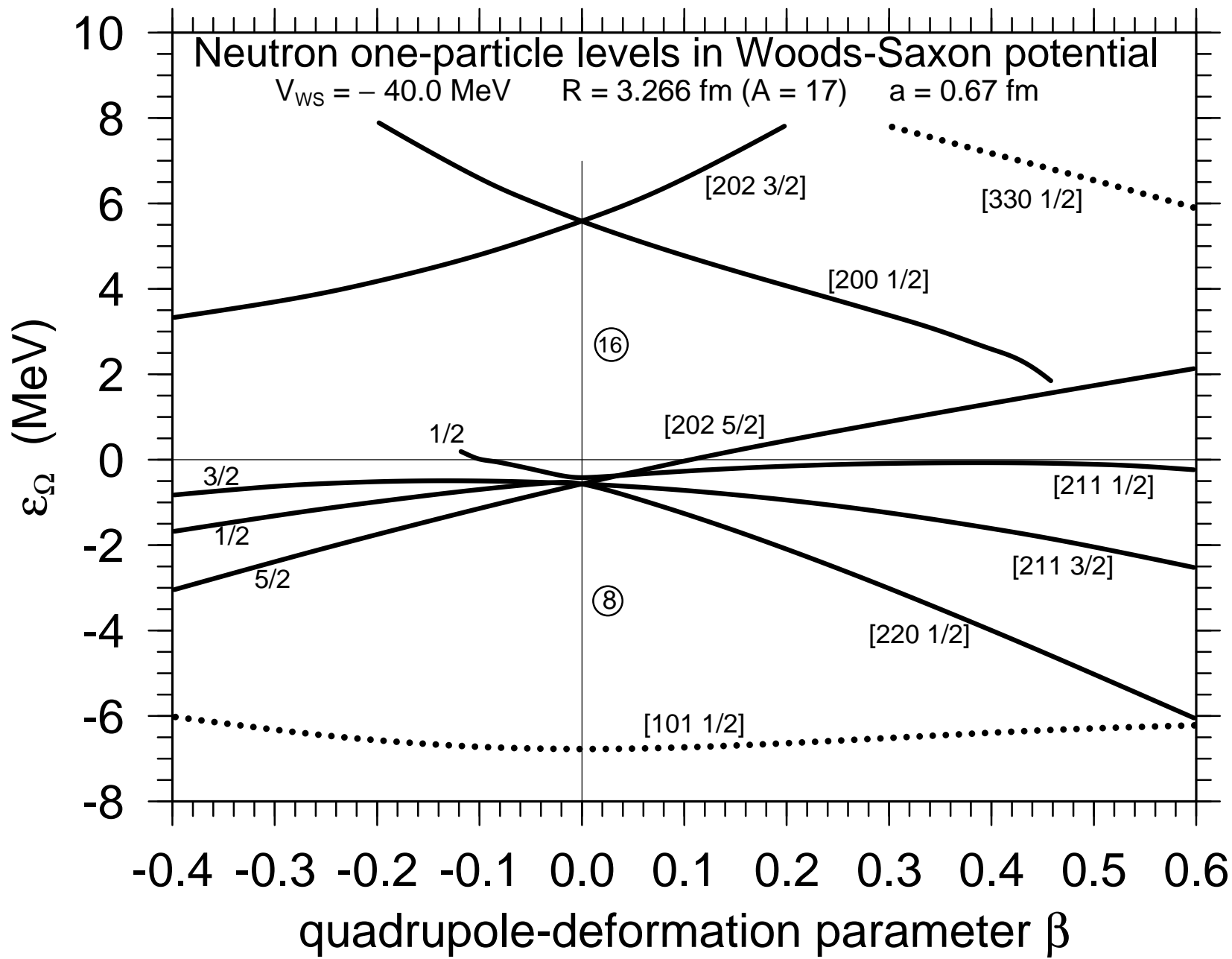
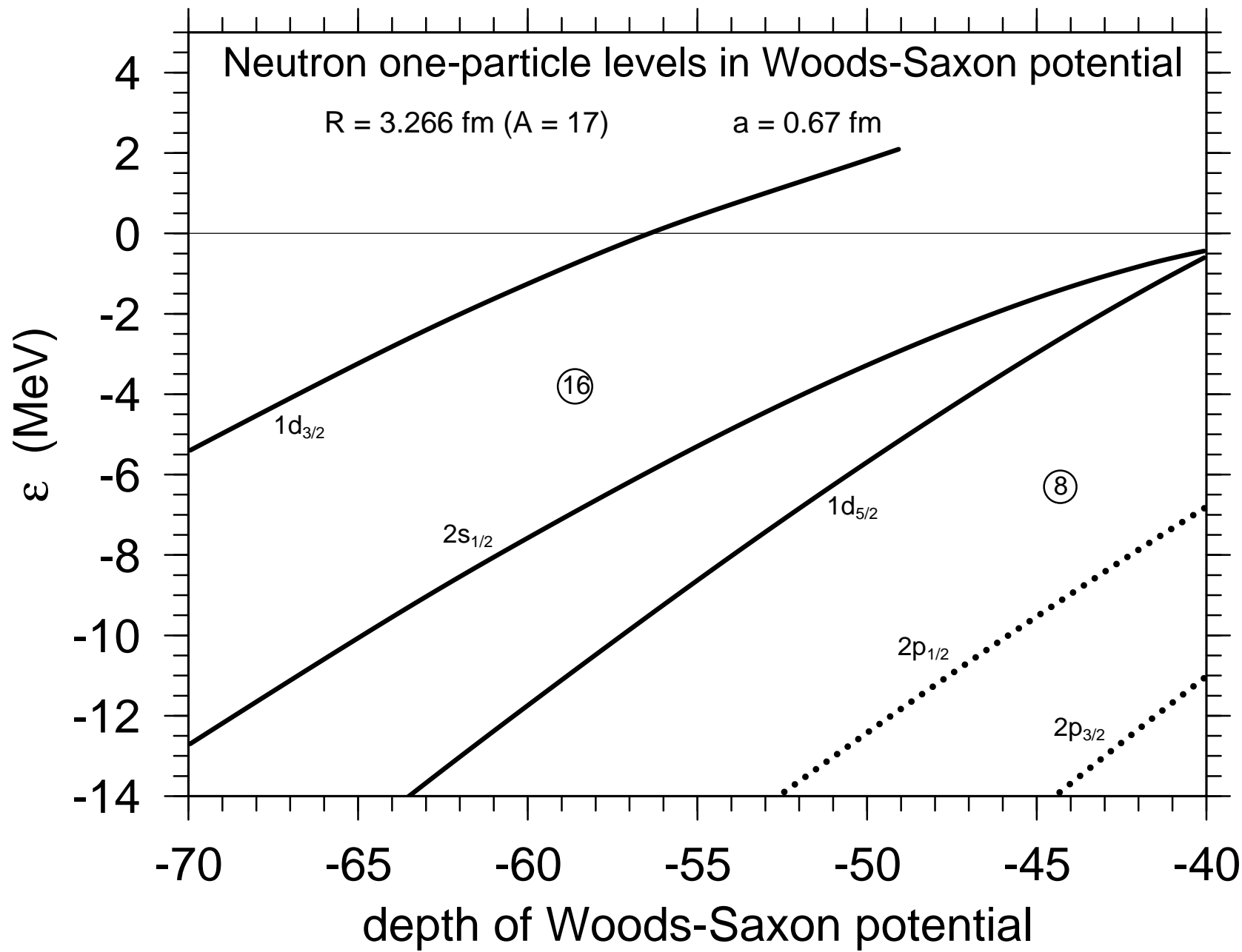
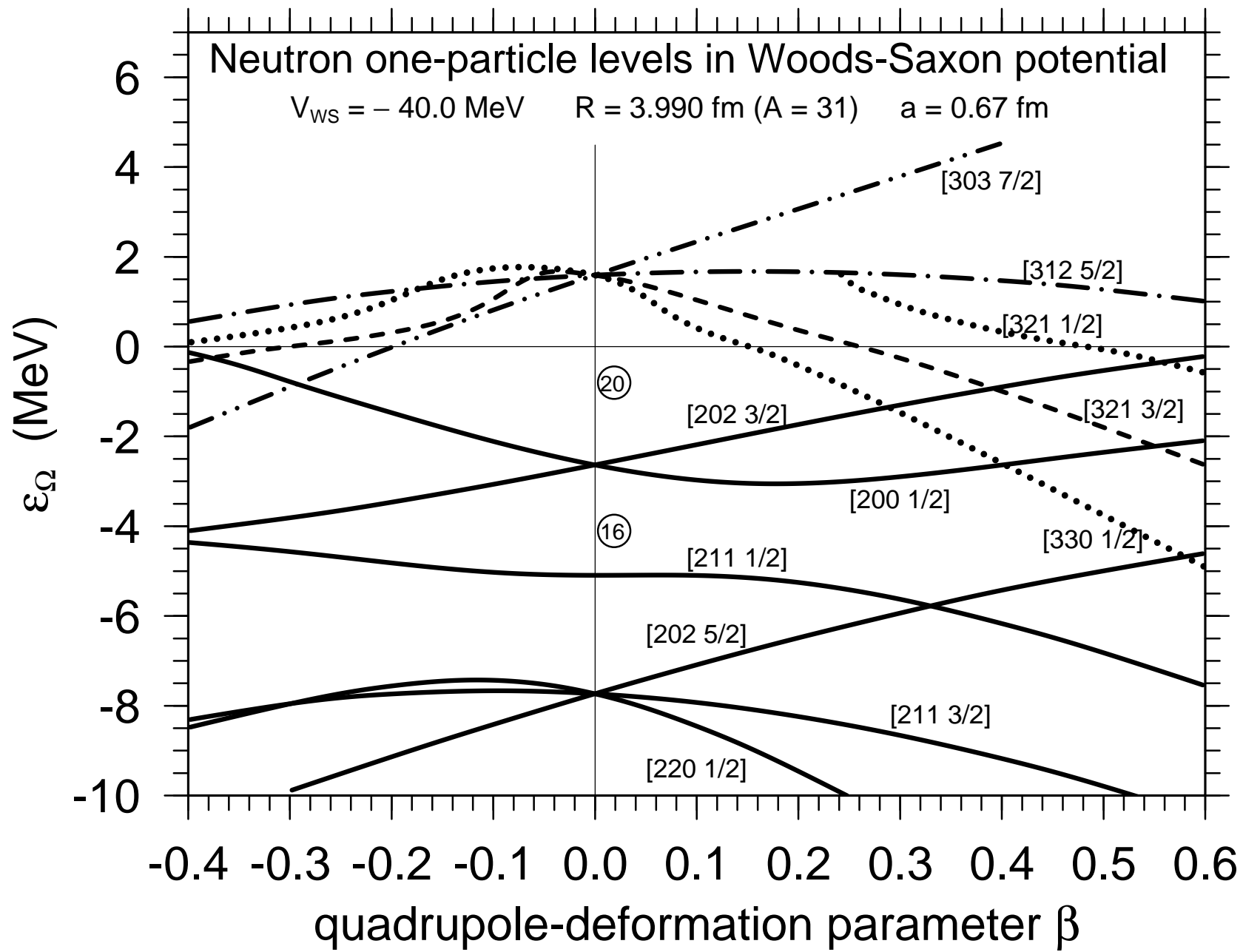
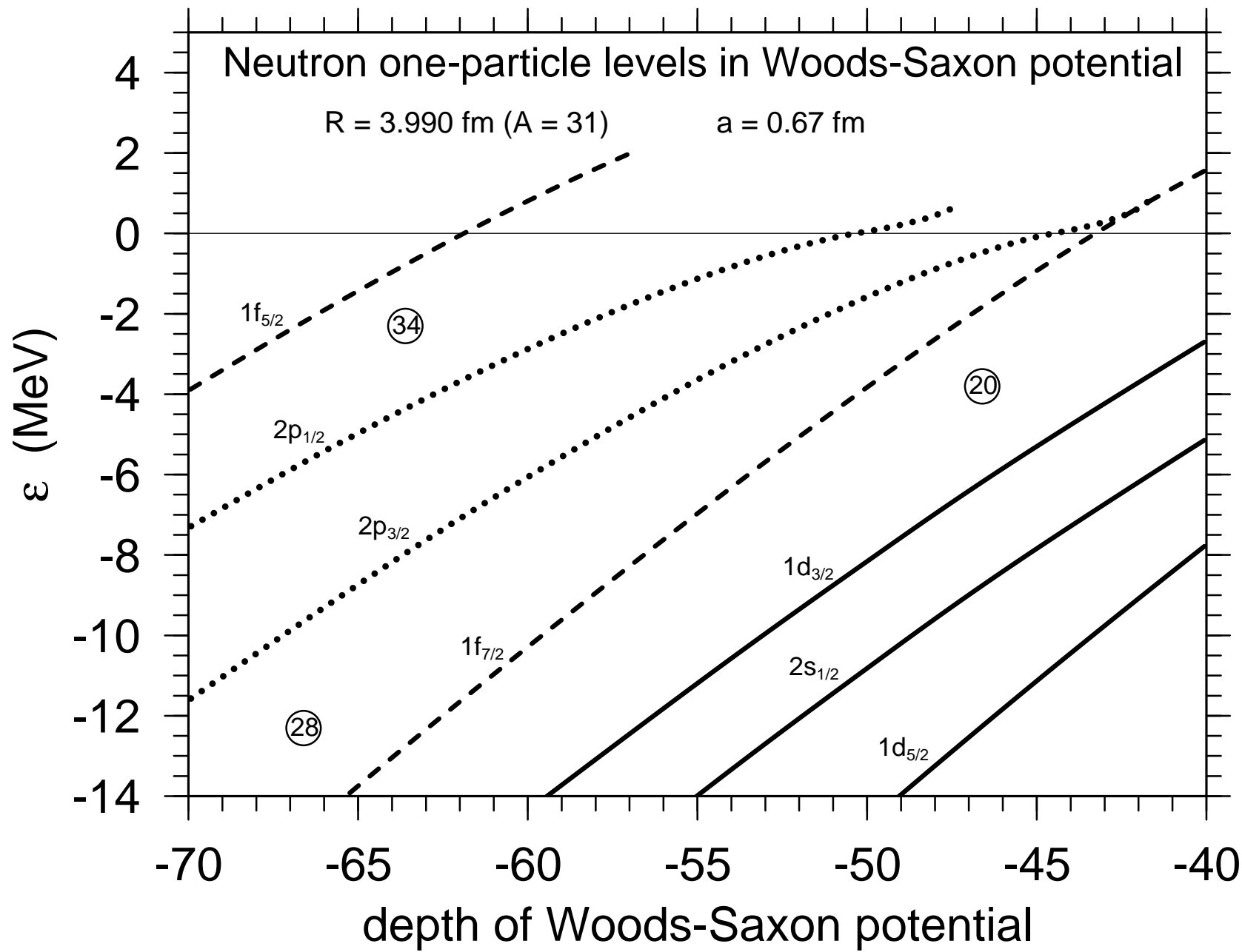
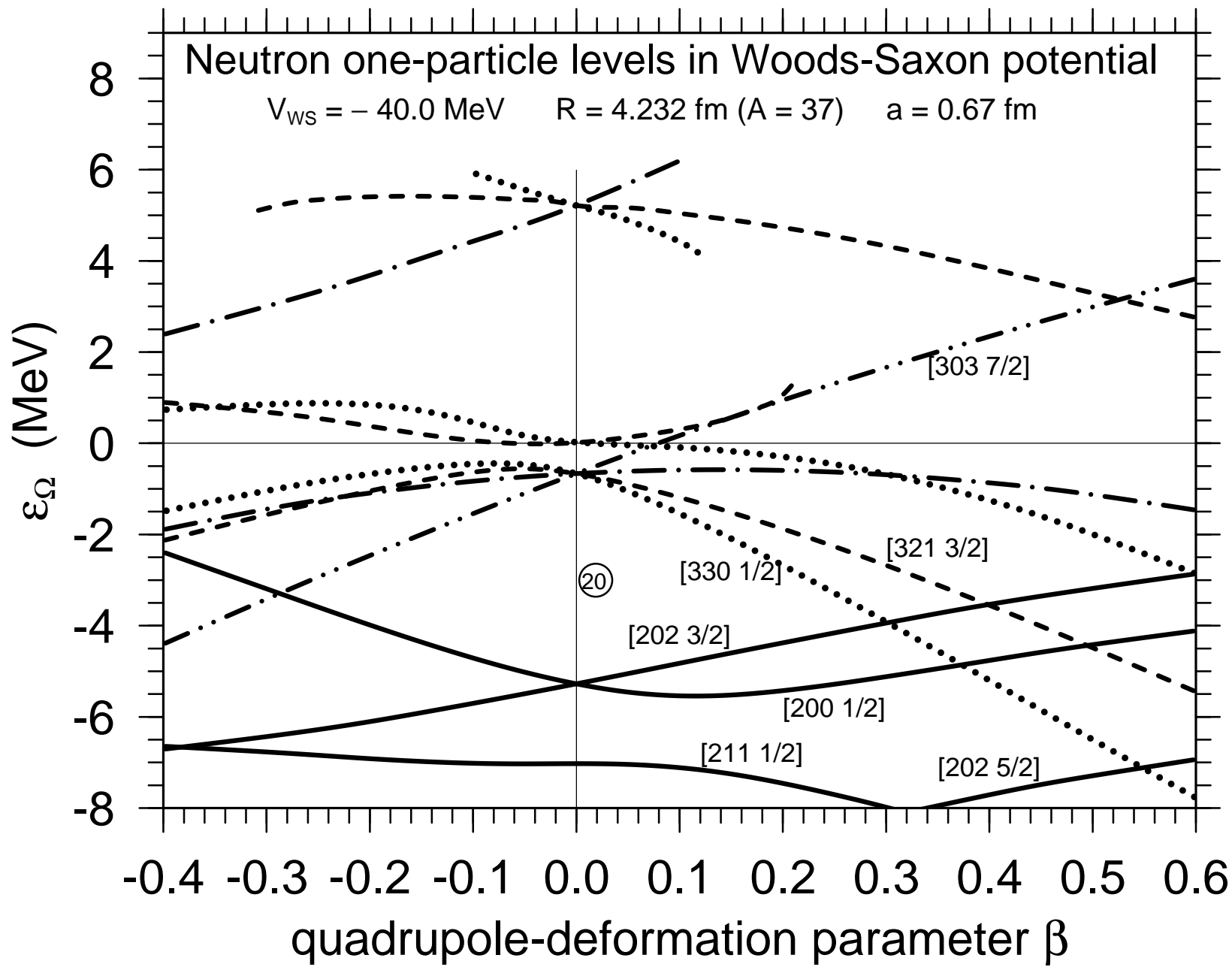
\end{document}